\documentclass[prl,aps,floats,amssymb,twocolumn,showpacs,preprintnumbers]{revtex4}

\usepackage{epsfig}

\newcommand{\fpi}{f_\pi}
\newcommand{\mpi}{m_\pi}
\newcommand{\chpt}{$\chi$PT}
\newcommand{\gev}{\, {\rm GeV}}
\newcommand{\mev}{\, {\rm MeV}}

\newcommand{\myfigsize}{\columnwidth}

\begin{document}

\preprint{ADP-03-112/T550}

\title{Physical Nucleon Properties from Lattice QCD}

\author{D.B.~Leinweber, A.W.~Thomas and R.D.~Young}
\affiliation{	Special Research Centre for the
                Subatomic Structure of Matter,
                and Department of Physics and Mathematical Physics,
                Adelaide University, Adelaide SA 5005,
                Australia}

\begin{abstract}
We demonstrate that the extremely accurate lattice QCD data for the
mass of the nucleon recently obtained by CP-PACS, combined with modern
chiral extrapolation techniques, leads to a value for the mass of the
physical nucleon which has a systematic error of less than one percent.

\end{abstract}


\maketitle



Hadronic physics presents fascinating theoretical challenges to the
understanding of strongly interacting systems in terms of their
fundamental degrees of freedom in QCD, quarks and gluons.  Traditional
perturbative methods in quantum field theory cannot account for the
highly non-trivial structure of the QCD vacuum -- which is dominated
by large scale quark and gluon correlations.  Lattice gauge theory
\cite{Wilson:1974sk} has so far provided the only rigorous method of
solving non-perturbative QCD.  We will show that recent progress
within the field \cite{AliKhan:2001tx}, together with advances in
effectively field theory (EFT) \cite{Young:2002ib}, now permit the
accurate, model-independent extraction of the physical properties of
hadrons from lattice QCD simulations, even though it is not yet
feasible to make calculations at the physical quark mass.

It is now possible to make extremely accurate calculations of the
masses of hadrons, such as the nucleon, with dynamical
fermions. Indeed the CP-PACS group has just reported data with a
precision of order 1\% \cite{AliKhan:2001tx}.  However, such precise
evaluations are limited to quark masses an order of magnitude larger
than those found in nature. In order to compare with experiment, which
is after all one of the main aims in the field, it is therefore
necessary to extrapolate in quark mass.  It is well known that such an
extrapolation is complicated by the unavoidable non-analytic behaviour
in quark mass, which arises from Goldstone boson loops in QCD with
dynamically broken chiral symmetry~\cite{Pagels:1975se}.

Early work motivated by the important role of Goldstone bosons led to
the construction of chiral quark models~\cite{Thomas:1984kv}, which
incorporated this non-analytic behavior. An alternative, systematic
approach, designed to avoid reference to a model, involved the
construction of an effective field theory (EFT) to describe QCD at low
energy \cite{Weinberg:1979kz}. The application to baryons has
developed to the point where chiral perturbation theory (\chpt) is now
understood as a rigorous approach near the chiral limit
\cite{Gasser:1988rb,Bernard:1995dp}.

Because it is defined as an expansion in momenta and masses about the
chiral limit, \chpt\ provides an attractive approach to the problem of
quark mass extrapolation for lattice QCD. The advantages of
formulating \chpt\ with a physical regularisation scheme, as opposed
to the commonly implemented dimensional regularisation, have been
demonstrated by Donoghue {\it et al.} \cite{Donoghue:1998bs}.  Early
implementations of a finite-range regulator (FRR) to evaluate chiral
loop integrals in \chpt\ suggested that, in the context of the
extrapolation of lattice data from relatively large quark masses, FRR
provides a more reliable procedure~\cite{Leinweber:1999ig}. There has
been considerable debate on whether current lattice data is within the
scope of dimensionally-regularised \chpt\ or whether the form of the
FRR chosen introduces significant model-dependence
\cite{Bernard:2002yk}.  However, this issue has now been
resolved~\cite{Young:2002ib} through a recent detailed study of
numerous regularisation schemes in \chpt, both dimensional and
FRR. This study quantified the applicable range of the EFT
\cite{Young:2002ib} and established that all of the FRR considered
provided equivalent results over the range $m_\pi^2 \lesssim
0.8\gev^2$.

Here we demonstrate that by adopting the FRR formulation of \chpt\ one
can improve its convergence properties to the point where
the chiral extrapolation to the physical quark mass can be
carried out in a model-independent way with a systematic
uncertainty of less than one percent. Given this remarkable result, 
the current errors on extrapolated quantities are dominated by the
statistical errors arising from the large extrapolation distance --   
the lightest simulated pion mass being typically 
$\mpi^2\simeq 0.27\, {\rm GeV}^2$, in comparison with 
the physical value, $0.02\, {\rm GeV}^2$. 

In the usual formulation of effective field theory the nucleon mass as
a function of the pion mass (given that $m_\pi^2 \propto m_q$
\cite{Gell-Mann:1968rz}) has the formal expansion:
\begin{eqnarray}
M_N &=& a_0 + a_2 m_\pi^2 + a_4 m_\pi^4 + a_6 m_\pi^6 + \ldots \nonumber \\
    & & + \sigma_{{\rm N} \pi} + \sigma_{\Delta \pi}.
\label{eq:formal}
\end{eqnarray}
In principle the coefficients, $a_n$, can be expressed in terms of the
parameters of the underlying effective Lagrangian to a given order of
chiral perturbation theory. In practice, for current applications to
lattice QCD, the parameters must be determined by fitting to the
lattice results themselves. The additional terms, $\sigma_{{\rm N}
\pi}$ and $\sigma_{\Delta \pi}$, are the two self-energy terms
involving the (Goldstone) pion which yield the leading and
next-to-leading non-analytic behaviour of $M_N$. As these terms
involve the coupling constants in the chiral limit, which are
essentially model-independent \cite{Li:1971vr}, the only additional
complication they add is that the ultra-violet behaviour of the loop
integrals must be regulated in some way.

Traditionally, one uses dimensional regularisation which (after
infinite renormalisation of all even powers of $m_\pi$) leaves only
the non-analytic terms -- $c_{\rm LNA} m_\pi^3$ and $c_{\rm NLNA}
m_\pi^4 \ln (m_\pi/\mu)$, respectively. Within dimensional
regularisation one then arrives at a truncated power series for the
chiral expansion,
\begin{eqnarray}
M_N &=& c_0 + c_2 m_\pi^2 + c_{\rm LNA} m_\pi^3 + c_4 m_\pi^4 \nonumber \\
    & & + c_{\rm NLNA} m_\pi^4 \ln \frac{m_\pi}{\mu}
	+ c_6 m_\pi^6 + \ldots \, ,
\label{eq:drexp}
\end{eqnarray}
where the bare parameters, $a_i$, have been replaced by the finite,
renormalised coefficients, $c_i$. Through the chiral logarithm one has
an additional mass scale, $\mu$, but the dependence on this is
eliminated by matching $c_4$ to ``data'' (in this case lattice
QCD). Provided the series expansion in Eq.~(\ref{eq:drexp}) is
convergent over the range of values of $m_\pi$ where the lattice data
exists, one can then use Eq.~(\ref{eq:drexp}) to evaluate $M_N$ at the
physical pion mass.  Unfortunately there is considerable evidence that
this series is not sufficiently
convergent~\cite{Young:2002ib,Bernard:2002yk,Thomas:2002sj,SVW,Hatsuda:1990tt}.

The physical reason for the lack of convergence seems to be that,
apart from the usual symmetry breaking scale in \chpt , $\Lambda_{\chi
{\rm SB}} \sim 4\pi f_\pi \sim $ 1 GeV, there is another mass scale in
the problem which does not appear explicitly in a dimensionally
regularised theory. This additional scale, $\Lambda$, is the inverse
of the size of the nucleon, the source of the pion cloud
\cite{Detmold:2001hq}. Of course, this scale must appear implicitly in
the correct effective field theory as a correlation between the
coefficients $c_n$. However, since we want to access pion masses above
0.5 GeV and the nucleon size of 0.5--1.0 fm suggests $\Lambda \sim
0.4$ GeV, it is clear that $m_\pi/\Lambda > 1$ and the naive series
should not be expected to converge over the range of pion masses of
interest. This problem can be overcome through the use of a FRR
\cite{Young:2002ib,Donoghue:1998bs}.

Clearly, in extracting information from lattice data for physical
hadrons one must avoid model dependence. In line with the implicit
$\mu$-dependence of the coefficients in the familiar dimensionally
regulated \chpt , in the case of a FRR the systematic expansion of the
nucleon mass is:
\begin{eqnarray}
M_N &=& a_0^\Lambda + a_2^\Lambda m_\pi^2 + a_4^\Lambda m_\pi^4 \nonumber \\
    & & + \sigma_{{\rm N} \pi}(m_\pi, \Lambda) + 
	\sigma_{\Delta \pi}(m_\pi, \Lambda)  ,
\label{eq:finite}
\end{eqnarray}
where the dependence on the {\it shape} of the regulator is
implicit. As for the mass scale, $\mu$, entering the usual chiral
logarithms, the dependence on the value of $\Lambda$ and the choice of
regulator is once again eliminated, to the order of the series
expansion, by fitting the coefficients, $a_n^\Lambda$, to lattice
data. The clear indication of success in eliminating model dependence
and hence having found a suitable regularisation method, is that the
higher order coefficients (for the best fit value of $\Lambda$) should
be small and that the result of the extrapolation should be
insensitive to the shape of the ultraviolet regulator.

In order to investigate the model dependence associated with the
self-energies, several regulators are considered. We evaluate 
the loop integrals in the heavy baryon limit  
\begin{equation}
\sigma_{BB'}^\pi =  - \frac{3}{16\pi^2 \fpi^2} G_{BB'}
\int_0^\infty dk \frac{k^4 u^2(k)}
{\omega(k) ( \omega_{BB'} + \omega(k) )},
\label{eq:fullSEpi}
\end{equation}
taking $u(k)$ to be either a sharp cut-off, $\theta(\Lambda - k)$, a
dipole, $(1 + k^2/\Lambda^2)^{-2}$, a monopole, $(1 +
k^2/\Lambda^2)^{-1}$ or finally a Gaussian, $\exp(-k^2/\Lambda^2)$.
We note that the dipole is the most physical, in that the observed
axial form factor of the nucleon has this shape
\cite{Thomas:2001kw}. In Eq.~(\ref{eq:fullSEpi}) for $B=B'=N$ we have
$G_{NN} = g_A^2$ (with $g_A = 1.26$), while $G_{N \Delta}=32
g_A^2/25$, the SU(6) value. In addition, $\omega
(k)=\sqrt{k^2+\mpi^2}$, $\omega_{NN}=0$ and $\omega_{N \Delta} = 292$
MeV, the physical $\Delta$-$N$ mass splitting.  Clearly the four
choices of regulator have very different shapes, with the only common
feature being that they suppress the integrand for momenta greater
than $\Lambda$.

In addition to the FRR expansions of Eq.~(\ref{eq:finite}) and the
standard dimensionally regulated expansion of Eq.~(\ref{eq:drexp}), we
also consider the case where Eq.~(\ref{eq:drexp}) is modified to
maintain the square-root branch-cut at $\mpi=\omega_{N \Delta}$
\cite{Young:2002ib}.  One can now compare the expansion about the
chiral limit for these six different regularisation schemes in order
to assess their rate of convergence.
It turns out that all the FRR expansions precisely describe the
dimensional regularization expansion over the range $m_\pi^2 \in
(0,0.7)$ GeV$^2$. Furthermore, the smooth, FRR formulations are
consistent with each other, for the renormalized chiral coefficients,
$c_{0,2,4}$, to an extraordinarily precise level~\cite{Young:2002ib}.
This ensures a reliable extrapolation to the regime of physical quark
masses.

We now exploit the accuracy of the most recent CP-PACS data
\cite{AliKhan:2001tx} by using it to determine either the first four
unknown parameters, $c_0-c_6$, appearing in Eq.~(\ref{eq:drexp})
(i.e. using dimensional regularisation, with and without the $\Delta
\rightarrow N \pi$ branch point) or the four parameters $a_0,a_2,a_4$
and $\Lambda$ in Eq.~(\ref{eq:finite}), for each choice of
ultra-violet regulator \cite{r0scale}. It is only the extreme accuracy
of the data which makes the determination of as many as four
parameters possible.
\begin{figure}[t]
\begin{center}
\epsfig{file=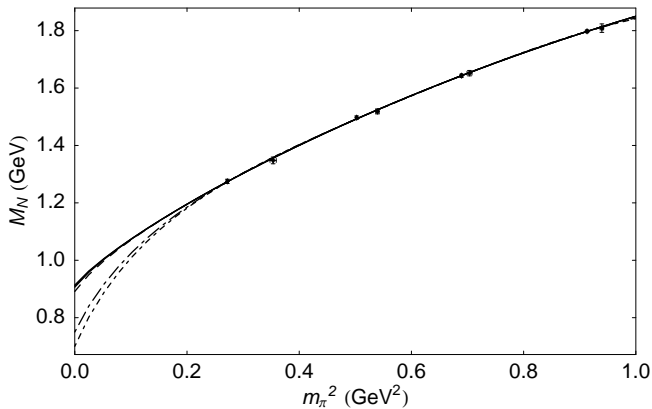,angle=0,width=\myfigsize}
\caption{Fits to lattice data for various ultra-violet regulators.
The monopole, dipole and Gaussian cases are depicted by solid lines, {\bf
indistinguishable} on this plot, the sharp cutoff is shown by the
dashed curve. The dimensional regularised forms are illustrated by the
dash-dot curves, with the correct branch point corresponding to the
higher curve. Lattice data is from
Ref.~\protect\cite{AliKhan:2001tx}.}
\label{fig:latfit}
\end{center}
\end{figure}
Figure~\ref{fig:latfit} shows the resulting fits to the lattice data
over the range $m_\pi^2 \in (0,1.0)\gev^2$, with the corresponding
parameters given in Table~\ref{tab:afit}. It is remarkable that three
of the four curves based on Eq.~(\ref{eq:finite}), the dipole, monopole
and Gaussian cases, are indistinguishable on this plot. Furthermore, we see
from Table~\ref{tab:afit} that the coefficient of $m_\pi^4$ in all of
those cases is quite small -- an order of magnitude smaller than the
higher order coefficients obtained using Eq.~(\ref{eq:drexp}), which
also oscillate in sign. This indicates that the residual series,
involving $a_{0,2,4}$, is converging when the chiral non-analytic behaviour is
evaluated with a FRR.
\begin{table}
\begin{center}
\begin{ruledtabular}
\begin{tabular}{lrrrrr}
Regulator	& $a_0$    & $a_2$    & $a_4$    & $a_6$    & $\Lambda$ \\
\hline
Dim.~Reg.	& $0.700$  & $5.10$   & $3.87$   & $-2.35$  & --        \\
Dim.~Reg.~(BP)	& $0.691$  & $5.32$   & $7.65$   & $-1.36$  & --        \\
Sharp Cutoff	& $1.06$   & $1.06$   & $-0.249$ & --       & $0.440$   \\
Monopole	& $1.38$   & $0.950$  & $-0.217$ & --       & $0.443$   \\
Dipole		& $1.15$   & $0.998$  & $-0.227$ & --       & $0.732$   \\
Gaussian	& $1.11$   & $1.02$   & $-0.234$ & --       & $0.593$
\end{tabular}
\end{ruledtabular}
\caption{Bare, unrenormalised, parameters extracted from the fits to lattice
data displayed in Figure~\ref{fig:latfit}. (All quantities are in
units of appropriate powers of GeV and $\mu=1$ GeV 
in Eq.~(\protect\ref{eq:drexp}).)
\label{tab:afit}}
\end{center}
\end{table}

As explained by Donoghue {\it et al.}  \cite{Donoghue:1998bs}, one can
combine the order $m_\pi^{0,2,4}$ terms from the self-energies with
the ``bare'' expansion parameters, $a_{0,2,4}$, to obtain physically
meaningful renormalized coefficients.  These are shown in
Table~\ref{tab:cren}, in comparison with the corresponding
coefficients found using Eq.~(\ref{eq:drexp}). Details of this
renormalization procedure are given in Ref.~\cite{Young:2002ib}.
\begin{table}
\begin{center}
\begin{tabular}{lrrr}
\hline
\hline
Regulator\hspace*{9mm}	& $c_0$ & \hspace*{8mm}$c_2$ & \hspace*{8mm}$c_4$  \\
\hline
Dim.~Reg.	& $0.700$ & $5.10$ & $3.87$ \\
Dim.~Reg.~(BP)	& $0.750$ & $4.59$ & $6.43$ \\
Sharp cutoff	& $0.892$ & $3.16$ & $12.9$ \\
Monopole	& $0.914$ & $2.63$ & $26.1$ \\
Dipole		& $0.910$ & $2.72$ & $23.4$ \\
Gaussian	& $0.906$ & $2.79$ & $21.4$ \\
\hline
\hline
\end{tabular}
\caption{Renormalised expansion coefficients in the chiral limit
obtained from various regulator fits to lattice data. (All quantities
are in units of appropriate powers of GeV.)
\label{tab:cren}}
\end{center}
\end{table}
The degree of consistency between the best-fit values found using all
three smooth choices of FRR is remarkably good, whereas the large
numerical values and the change of sign in the coefficients found from
the fit using Eq.~(\ref{eq:drexp}) suggest that the latter is nowhere
near convergence. We can understand the problem very simply; it is not
possible to accurately reproduce the necessary $1/m_\pi^2$ behaviour
of the chiral loops (for $m_\pi > \Lambda$) with a 3rd order
polynomial in $m_\pi^2$. The final point to notice about the finite
range regulators is that all but the sharp cutoff naturally yield a
correction to the Goldberger-Treiman relation of approximately the
right size. As shown by Birse and McGovern~\cite{McGovern:1998tm},
this correction dominates the non-analytic term of order $m_\pi^5$
which only arises at two loops in dimensionally regulated \chpt . The
absence of the correct $m_\pi^5$ term in the case of the sharp cutoff
explains its slight deviation from the results of the other FRR.

It is clear that the use of an EFT with FRR enables one to make a
model-independent extrapolation of the nucleon mass as a function of
the quark mass. Although minimal deviation is seen between the best
fit curves, we need to determine how well these curves are in fact
constrained by the statistical uncertainties of the lattice data. As
all data points are statistically independent, the one-sigma deviation
from the best-fit curve is defined by the region for
$(\chi^2-\chi^2_{\rm min})/dof < 1$. We use a standard $\chi^2$
measure, weighted by the squared error of the simulated data point and
$\chi^2_{\rm min}$ corresponds to the optimum fit to the data.  To
preserve the infrared physics of EFT, we put a lower bound of
$300\mev$ (just above the $N-\Delta$ mass splitting) on the regulator
parameter, $\Lambda$.

\begin{figure}[t]
\begin{center}
\epsfig{file=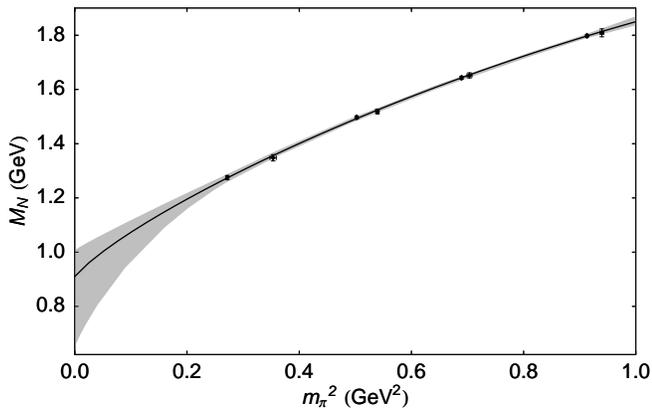,angle=0,width=\myfigsize}
\caption{Error analysis for the extraction of the nucleon mass using a 
dipole regulator. The shaded region corresponds to the region 
allowed within the present statistical errors.
\label{fig:diperr}}
\end{center}
\end{figure}

We show the one-sigma variation from the best-fit dipole curve by the
shaded region in Figure~\ref{fig:diperr}. The extrapolated values for
the nucleon mass are shown in Table~\ref{tab:physical}. It is clear
that the large lower-deviation on the extrapolated values are a result
of the regulator mass not being constrained from above. Alternatively,
for the FRR cases, one could fix $\Lambda$ and vary an additional
parameter, $a_6$, in the spirit of dimensional regularisation. In that
case the error bounds are symmetric and similar in magnitude to those
found for the dimensionally regulated cases reported in
Table~\ref{tab:physical}. The primary source of the large error band
is the large extrapolation distance.

We also quote the $\sigma$-commutator in Table~\ref{tab:physical} --
calculated as $m_\pi^2 \partial M_N/ \partial m_\pi^2$ at the physical
pion mass~\cite{Leinweber:2000sa}. Again the statistical errors are
large. In this case, as well as for the nucleon mass itself, it is
clear that a simulated point at a pion mass $\mpi^2\sim 0.1\gev^2$
would greatly reduce the statistical error in the extrapolation. We
note that our study has focused on the systematic errors associated
with chiral extrapolation, while those arising from finite lattice
spacing and volume~\cite{Young:2002cj,SVW} remain to be quantified.
\begin{table}
\begin{center}
\begin{tabular}{lrr}
\hline
\hline
Regulator\hspace*{9mm}	& $m_N$ & \hspace*{9mm} $\sigma_N$   \\
\hline
Dim.~Reg.	& $0.782^{+0.122}_{-0.122}$  & $73^{+15}_{-15}$ \\
Dim.~Reg.~(BP)	& $0.823^{+0.122}_{-0.122}$  & $65^{+15}_{-15}$ \\
Sharp cutoff	& $0.939^{+0.078}_{-0.219}$  & $41^{+35}_{-12}$  \\
Monopole	& $0.953^{+0.063}_{-0.219}$  & $35^{+36}_{- 8}$  \\
Dipole		& $0.951^{+0.081}_{-0.217}$  & $36^{+35}_{-13}$  \\
Gaussian	& $0.948^{+0.080}_{-0.216}$  & $37^{+35}_{-13}$  \\
\hline
\end{tabular}
\caption{Here we show the nucleon mass, $m_N$ (GeV), and the sigma
commutator, $\sigma_N$ (MeV), extrapolated to the physical pion
mass.
\label{tab:physical}}
\end{center}
\end{table}

To summarize, we have shown that the extremely precise dynamical
simulation data from CP-PACS permits us to determine four parameters
in the chiral extrapolation formulas, Eqs.~(\ref{eq:drexp}) and
(\ref{eq:finite}). Whereas the former (involving
dimensional-regularization) does not appear to be convergent over the
required mass range, the improved convergence properties of a finite
range regulator yield an excellent description of the data over the
full mass range, regardless of the functional form chosen for the
vertex regulator. Table~\ref{tab:physical} summarises the resulting
values of the nucleon mass and the $\sigma$-commutator evaluated at
the physical pion mass. The nucleon mass is constrained by the chiral
extrapolation technique based upon a FRR procedure to a systematic
error less than $1\%$. The $\sigma$-commutator is similarly
constrained to within a systematic error of $3\%$. The values obtained
are both consistent with earlier extractions from lattice data using a
similar technique \cite{Leinweber:2000sa} and, within the relatively
large statistical errors, in agreement with the experimental values
\cite{Gasser:1991ce}. We note also that the systematic uncertainty in
the extraction of the low energy constant, $c_0$, is less than 1\%,
while for $c_2$ it is at the level of a few percent. With the issue of
model independence resolved, there is an urgent need for high
precision lattice QCD simulations at $m_\pi^2 \sim 0.1$ GeV$^2$ in
order to reduce the present statistical error on the
extrapolation. Such simulations should be feasible with the new
generation of lattice QCD computers currently under construction.

\acknowledgments We thank M.~Birse, J.~McGovern and S.V.~Wright for
helpful conversations. This work was supported by the Australian
Research Council.


\end{document}